# Artificial Intelligence – from Idea to Implementation. How Can AI Reshape the Education Landscape?

Cătălin Vrabie

**Abstract**
This introductory chapter provides an overview of the evolution and impact of Artificial Intelligence (AI) technologies in today's society. Beginning with a historical context while exploring a few general definitions of AI, the author provides a timeline of the used technologies, highlighting its periods of stagnation, commonly referred to as "AI winters," and the subsequent resurgence fueled by relentless enthusiasm and investment. The narrative then transitions to focus on the transformative effects of AI on society at large, with a particular emphasis on educational applications. Through examples, the paper shows how AI technologies have moved from theoretical constructs to practical tools that are reshaping pedagogical approaches and student engagement. The essay concludes by discussing the prospects of AI in education, emphasizing the need for a balanced approach that considers both technological advancements and societal implications.

## Introduction

We have learned from our mistakes throughout history to adapt to a hostile environment. We have also learned how to refine the tools we develop to guarantee that they continue to be beneficial. For example, after inventing fire, which often got out of control, we went on to invent fire extinguishers, fire alarms, and develop fire services. Similarly, the invention of gunpowder and firearms led to the creation of bulletproof vests and armor-plated vehicles and the development of guard and protection services. The invention of cars was followed by the introduction of seat belts, airbags, and, more recently, self-driving automobiles. It is safe to say that technology is an expression of human will. Through technological advancements, we seek to extend our control over various aspects of our environment – be it distance, nature, or even interpersonal dynamics. Each of the tools we developed possesses the power to influence our perspectives and shape the future (Vrabie & Eduard, 2018; Vrabie, 2016). For example, farming tools have revolutionized agricultural practices, and lab instruments have opened new frontiers for scientists. Books, maps, and similar devices, often called "intellectual technologies" (Goody & Bell, 1975), have expanded our world understanding. These last ones, in particular, have had the most significant impact on society as we know it.

Today, computers and the Internet are among our most personal tools – when IBM decided to market its first microcomputer (1981) as a PC (Personal Computer), it was not a coincidence (The New York Times, 1981). They serve as platforms for self-expression and play a crucial role in shaping our personal and public identities and fostering relationships within our communities (Lazar, Zbuchea, & Pînzaru, 2023).

Artificial Intelligence (AI) represents a relatively new category of technological tools, referred to as "smart technologies" (Campbell, 2023; NanoWerk, 2023) and they are primarily built upon intellectual ones. However, unlike them, smart technologies aim not only to amplify or complement our native capabilities but to augment them – and sometimes, even to replace them.

**In search of a definition**

The term "Artificial Intelligence" is notably challenging to define. At times, it is employed to describe tasks difficult for computers to execute, such as engaging in verbal dialogue, as opposed to functions they already excel at, like spreadsheet calculus and computation. Recently, amid the growing hype surrounding AI, IT companies have started to label their latest products as "innovative" by using the term to emphasize virtually any capability computers possess, including, for example, the operation of traditional databases.

The concept we aim to explore throughout this chapter is often mentioned in the media and casual discussions, giving the impression that it is well-understood. However, it eludes a single, indisputable definition. Numerous competing definitions exist, encompassing logic and understanding, planning ability, emotional development and awareness, creativity, problem-solving, and perhaps most importantly, the ability to learn.

The psychologist and Harvard professor Howard Gardner focused on an individual's aptitude for finding solutions in various situations during his career. He defined human Intelligence as the ability to solve problems within one or more cultural contexts (Gardner, 1983; Vidu, Zbuchea, & Pinzaru, 2021). In stark contrast, the term "artificial" is relatively straightforward, denoting something not naturally occurring. This creates an apparent oxymoron[1] when paired with "intelligence," which is a natural phenomenon.

Informally, many use "artificial intelligence" to refer to tasks challenging for computers, such as understanding spoken language or optimizing routes via driving apps, instead of tasks where they excel, like high-precision calculations.

A crucial distinction exists between "narrow AI" and "general AI." Narrow AI involves computing systems designed to solve specific tasks within a bounded framework, like chess or Go-playing systems (Gregory Unruh & David Kiron, 2017). General AI, on the other hand, refers to systems capable of solving diverse problems autonomously, much like humans. As of this writing, all AI applications are in the realm of narrow AI. While general AI remains a subject of intense scientific interest, its successful implementation appears to be years, if not decades, away.

We'll begin with a straightforward definition of AI: machines that exhibit seemingly intelligent behavior[2] (Vidu, Zbuchea, Mocanu, & Pinzaru, 2020). While this catchy definition aptly encapsulates the term's contribution to IT, the reality is far more complex. To elaborate, I will provide background information on the history of AI, which will coalesce into a comprehensive definition of the concept under study.

---

[1] This idea has engendered considerable scholarly discourse across various scientific disciplines. However, the present article does not aim to provide a definitive resolution to these nomenclatural debates; rather, it concentrates on substantive elements of the subject matter.

[2] From a human perspective.

**Hello, world!**[3]

Modern discourse on the potential of intelligent systems can be traced back to as early as 1950 when Alan Turing published his seminal paper, "Computing Machinery and Intelligence" (Turing, 1950). Even then, researchers were exploring methods of automatic computation that could transform various aspects of life – from performing complex arithmetic to playing chess. Alan Turing was aware that defining Intelligence – especially in the context of machines – was a difficult task. Therefore, he introduced what became known as the "Turing Test." According to this test, if a human could not determine whether they were chatting with a computer or another human within five minutes, the computer would have passed the test and thus be deemed intelligent. At that time, conducting this dialogue verbally was not feasible, so it was carried out in written format[4].

Turing's primary aim was not to test Intelligence *per se* but rather to address the various objections people had about the feasibility of associating Intelligence with machines. In his paper, he explored a wide range of arguments. He cited the work of Charles Babbage, who attempted to build the Analytical Engine between 1828 and 1839. Also, he mentioned Countess Ada Lovelace, often considered the first programmer, who stated in 1842 that "the Analytical Engine has no pretensions whatever to originate anything. It can do whatever we know how to order it to perform." However, he disagreed by suggesting that we can "order" the machine to be original by programming it to produce answers we cannot predict (Turing, 1950)… Turing's core assertion was that there is no reason to doubt that machines will eventually achieve thought.

---

[3] This formula is conventionally employed as the initial output when executing a newly developed computer program. This practice serves as a humorous ritual among software developers, symbolizing the 'birth' of the application.

[4] The technology was called teletype (TTY).

Over a decade after the publication of "Computing Machinery and Intelligence," researchers like John McCarthy, Marvin Minsky, Allen Newell, Herb Simon, Nathaniel Rochester, and Claude Shannon began establishing laboratories focused on computer science. This marks the birth of artificial intelligence as a formal field, specifically in 1955, with "A proposal for the Dartmouth Summer Research Project on Artificial Intelligence," dated August 31 (McCarthy, Minsky, Rochester, & Shannon, 1955). This was the first academic initiative aimed at investigating the subject in depth. The outcome of this initial effort was a strategic breakdown of tasks. In essence, the participants outlined an agenda for initiatives that would span not just years but, ultimately, decades.

- First on their agenda was the development of rational machines capable of performing tasks similar to humans – such as playing chess, solving algebraic problems, proving geometric theorems, and even accurately diagnosing diseases. These tasks are generally perceived as problems that require reasoned solutions.
- Another key focus was enabling computers to understand the world around them. This entailed the creation of programming languages that could encode the existence of various elements, allowing machines to comprehend them.
- The third priority was to equip computers with the ability to navigate and plan within the world. This involved understanding how to move from one point to another, assessing routes, and determining the least risky options.

- Understanding and replicating human language, whether written or spoken, also emerged as a significant area of focus. The challenge lay in teaching machines to understand sentence construction and contextual meaning despite lacking human emotions and thoughts. The ultimate goal was to "teach computers as much as possible about what all these things mean."
- Lastly, perception – how we see, hear, and react to our environment – was identified as crucial for activating machine intelligence. This required the development of sensors and algorithms that could generate feedback based on sensory input. Initial experiments employed video cameras for observation and mechanical arms for task execution based on what was "seen."

The aspiration was that foundational elements for developing artificial intelligence could be established by enabling computers to perform the aforementioned tasks successfully. This included aspects like emotional intelligence, creative reasoning, intuition, and more. The goal was not to hard-code[5] human-like behavior into computers but to allow them to develop or acquire these capabilities through experiential learning, akin to the learning processes in children.

John McCarthy dedicated his career to leveraging mathematical tools to advance computer science. He invented Lisp in 1958, making it the second-oldest programming language after Fortran[6]. On the other hand, Allen Newell and Herb Simon concentrated on understanding human thought processes. They developed systems that solved puzzles and simple problems in ways they believed mirrored human cognition. Marvin Minsky emphasized the importance of context, arguing that computers should be programmed to mimic either human intuition or mathematical computation. He posited that relying solely on either approach would be insufficient for a comprehensive understanding of machine intelligence, a viewpoint he elaborated on in his 1960 paper, "Steps Toward Artificial Intelligence" (Minsky, 1960).

---

[5] A term in programmers' jargon indicating that the response of a machine is built-in and not the result of a process of analysis and/or computation.

In the early 1960s, a software industry pioneer, James Slaglerote, developed a program as part of his doctoral thesis that integrated symbolic expressions commonly used in mathematics (Slagle, 1961). The program aimed to emulate the problem-solving approaches of students new to mathematics. Remarkably effective, the program not only managed to solve routine calculations but also opened up new avenues of exploration. Due to its success, it is often cited as a flagship program of the first wave of AI. James Slagle's pioneering work was swiftly followed by other advancements, and by 1970, programs could interpret drawings, learn from examples, and even construct complex geometric structures like fractals and architectural elements.

In Romania during that time, the Polytechnic University of Timişoara developed the MECIPT-1 computer[7], which was instrumental in designing the dome of the Bucharest exhibition pavilion (Romexpo) and the Vidraru Dam on the Argeş River in 1965 (Baltac, 2017; UPB, 2021).

---

[6] Which is only a year older.
[7] The acronym stands for Electronic Computing Machine of the Timisoara Polytechnic Institute [Maşina Electronică de Calcul, Institutul Politehnic Timişoara].

While trying to fulfill the fourth priority, it's worth noting that even in those early days, computers could respond to queries like modern-day voice assistants like Siri and Alexa. Romania saw its first automatic computer translation from English to Romanian in 1962[8], marking a national debut in what we now refer to as Natural Language Processing (NLP).

These historical efforts impart a crucial lesson: if we accurately represent cognitive mechanisms, we come significantly closer to understanding and developing AI. The idea of representation is central to defining what Artificial Intelligence entails. In essence, AI involves models of thinking, perception, and action (Winston P. H., 1990). This naturally raises the question: what constitutes a model? A model is a construct we develop to approximate reality with some degree of accuracy. Models help us understand, explain, and even predict and control various outcomes, encapsulating the essence of scientific inquiry.

The importance of representation cannot be overstated. It serves as the foundation for models of thinking. In broad terms, representations are conventions used to describe real-world situations. For example, in a history museum, lead soldiers symbolize frontline troops; in mathematics, operators like "+" and "-" denote specific operations; in biology, a representation might involve listing the key characteristics of a cell; and maps serve as excellent examples of spatial representations.

Therefore, we can further refine the definition of Artificial Intelligence: AI involves representations that underpin thinking, perception, and action models. Moreover, the actual value of a representation lies in its ability to expose constraints, which in turn enable specific modes of operation. While this may sound complex – and indeed it is – the complexity escalates when we consider that these models must be organized within an overarching system capable of implementing them. This results in a highly intricate and complex architecture.

---

[8] The translated phrase was: "You explain the development of science and we help describe the examples". Compared to what was happening overseas in this field, where unpaid efforts to translate from Russian into English and vice versa (the Cold War was in full swing, hence the need) led to the withdrawal of funding, Romania was in in the middle of a technological boom.

Artificial Intelligence can thus be defined as an architectural system that implements methods enabled by constraints exposed through representations, which support models of thinking, perception, and action (Winston, 1990). Importantly, this definition encompasses not just the act of doing whatever it is meant to do but also the capacity for learning how to do it (Minsky, 2006) – each component of this definition is crucial.

The technology that fueled the second wave of AI emerged around fifteen years after the first wave. In the mid-1970s, Edward Shortliffe developed the MYCIN[9] system for diagnosing medical conditions. Much like James Slagle's program, MYCIN achieved remarkable success, diagnosing certain conditions nearly as accurately as a general practitioner. This second wave focused on rule-based systems for representing knowledge, and it was propelled by the rise of the so-called "expert systems." These systems dominated AI research for the subsequent decade, while rule-based approaches remain a vital part of the AI toolkit.

By the mid-1980s, the lofty promises of AI began to seem unattainable, leading to a period commonly referred to as the "AI winter." Few software programs or applications succeeded in supplanting human experts.

---

[9] The name is given by the suffix of antibiotic-type drugs.

Many startups aiming to do so failed, and the term "AI" became synonymous with failed investments in the stock market. Despite this "winter," there was a gradual yet consistent advancement in knowledge, paving the way for new applications. Enthusiasts persevered, undeterred by social, political, or financial pressures, even though AI largely remained under the radar until its third wave.

In February 1996, world chess champion Garry Kasparov, who had held the title for over a decade, accepted a challenge to play a six-game match against IBM's Deep Blue, the most powerful computer of its time. Kasparov won 4–2. However, a May 1997 rematch featuring an upgraded Deep Blue resulted in a 3½-2½ victory for the machine. This win, although controversial, marked a historic moment in both AI and chess, which increasingly began to benefit from computational assistance (Garry & Mig, 2017).

The research and development in AI technologies continued smoothly. In February 2010, Siri was launched on the Apple App Store, allowing iPhone users to communicate verbally with their devices. In 2011, IBM's Watson defeated the human champion on "Jeopardy," signaling a sort of AI renaissance.

Spurred by these developments, tech giants like IBM, Google, Meta/Facebook, Amazon, Microsoft, and Apple have intensified their efforts in developing AI systems. The third wave has been further fueled by the abundance of online data and increasing computational power. This has facilitated statistical analysis, trend identification, and pattern recognition, leading to a resurgence in Machine Learning (ML)[10] and the growing use of Neural Networks (NN).

---

[10] A term first introduced long ago, in 1959, by IBM and further elaborated in 1974 by Paul Werbos in his Harvard PhD thesis.

Neural networks can self-learn. Specifically, using mathematical techniques, they can differentiate contextual elements by analyzing relevant data sets. Despite their potential, they initially received little attention until British programmer Geoffrey Hinton showcased impressive results in a 2012 image classification competition (Image net, 2012).

This resemblance between biological neurons and artificial neural networks has led to skepticism and critiques. Communities of individuals enjoy demonstrating how these networks, and computers in general, can be fooled. A recent example involved a Tesla car mistaking the moon for a yellow traffic light, causing the vehicle to slow down. The incident made the car owner publicly mock Tesla's AI capabilities on Twitter (JordanTeslaTech, 2021). What is often noticed in reports from techno-skeptics is a forced comparison between computational phenomena and similar occurrences in nature. Such comparisons are misleading. While there may be similarities, they are insufficient grounds for direct comparison. For instance, comparing an airplane's wings to a bird's wings is futile, even though both enable flight and are heavier than air… similarly, there will never be a race between a car and a runner, and so on.

Setting criticisms aside, it's worth examining the current state of machine learning and, more specifically, neural networks. While these networks perform impressively, they neither see nor think like humans – in fact, they don't think at all. Their algorithms are geared towards recognition, akin to training. They represent a significant achievement in the intersection of mathematics and computer engineering and constitute an important chapter in the AI narrative (LeCun, 2020). However, they are not the entire story. The current consensus suggests that we need more from machine learning than neural networks have provided. This subject continues to generate both excitement and concern, fueling ongoing debate.

Elon Musk (2014) once remarked that Artificial Intelligence is akin to "summoning the demon" and represents our greatest existential threat. While this statement is alarming, it's worth noting that Musk is known for his provocative tweets[11]. He is neither the first nor the last to express such concerns. Marvin Minsky himself warned in 1970 that once computers gain control, humanity might never regain it. He continued, saying, "If we're lucky, they might decide to keep us as pets." Also, Allen Newell and Herbert Simon[12] shared a similar sentiment, stating that symbol systems modeling human problem-solving capabilities could be a threat, and the public should be alerted.

However, Edward Feigenbaum offered a different perspective in 1982. In his book "The Handbook of Artificial Intelligence" (Barr & Feigenbaum, 1981), he argued that rule-based systems could mitigate such threats. He also incorrectly predicted that Japan would use computer systems to dominate the world economically. Around 1990, Rodney Brooks drew parallels between the evolution of robotics and the colonial expansion of the 14th and 15th centuries. He believed that once robots could be miniaturized to the size of insects, the rest would be straightforward. While we have made significant advancements in robotics, replicating Human Intelligence remains elusive. Interestingly enough, Elon Musk has recently invested heavily in AI, contradicting his earlier stance.

---

[11] The statement is, however, surprisingly similar to the "tickling the dragon's tail" made by physicists at Los Alamos during the Manhattan Project, the project that developed the atomic bombs that were later dropped on the Japanese cities of Hiroshima and Nagasaki.

[12] Nobel Laureate in Economics (1978) "for his pioneering research on decision-making in economic organizations" and known for the concept of "bounded rationality".

Currently, we are in a period of rapid development in neural networks. The work of researchers like Geoffrey Hinton, Yann LeCun, and Yoshua Bengio, who won the 2018 Turing Award for their groundbreaking image labeling program, has significantly fueled interest in this field[13].

**What's inside?!**

So far, this first chapter provided an overview of artificial intelligence. Next, it will briefly describe how society can benefit from this industry since we are still in the introductory chapter.

Many AI applications employ mathematical techniques like linear regression, optimization, and probabilistic reasoning – techniques that have been part of the mathematical toolkit for decades, if not centuries. However, AI is not just computational power. So, what's next? In the coming years, most AI researchers will focus on the cutting edge of machine learning, particularly in the area now known as Deep Learning (DL). It seems computing power and large databases – commonly called big data – remain the most crucial elements.

In this new information age, the assets that IT companies (and others) will protect are shifting from applications to data. Google has made its Deep Neural Networks system, TensorFlow, available to all interested parties. This ensures that new hires will be more proficient in using it than those at competing firms. However, like other major players in the global IT landscape, Google does not share the data they consider a competitive advantage.

---

[13] Awarded by the Association for Computing Machinery (ACM) and which is the equivalent of the Nobel Prize in Computer Science.

Today, we see processing systems with enormous power – unimaginable a decade ago – being offered to users free of charge by companies in the field. This has led to a surge in the number of people taking up the study of Artificial Intelligence, drawn by the successes of Deep Learning. However, on an AI timeline, we may be at a stage similar to where Traian Vuia was when he first flew in March 1906[14] – at that time, no one was contemplating drones capable of delivering packages, transporting passengers, or networks of artificial satellites orbiting the planet. It all began with an idea that a few enthusiasts invested time and effort into.

In the AI industry, the next frontier appears to be insight systems – those equipped with advanced skills. These systems should be capable of summarizing, conceptualizing, and even generating content. While we often read about the manifestation of self-awareness in various articles (Yalalov, 2023; The New York Times, 2023), it is still at such a rudimentary stage that science wouldn't even consider it hypothetical (OpenAI, 2023; IEEE Spectrum, 2023; Fast Company, 2023).

Terms like "intelligence," "creativity," and "consciousness" have broad meanings and current AI technologies only encapsulate a fraction of these. However, AI systems are defined by a range of specific objectives, methods, constraints, representations, and architectures; no single hardware or software solution can meet all these require-ments. Therefore, many successful AI applications either automate routine tasks or enable human-machine collaboration, achieving results that neither could accomplish alone. While Machine Learning / Deep Learning is often touted as the future of AI, it is not the complete representation of what AI can be. Moreover, trustworthy AI systems must be able to explain their actions. To achieve this, we need to develop systems that think and learn in ways similar to humans.

---

[14] Aviation is another area that has enjoyed spectacular growth. While the first flight in a self-propelled, heavier-than-air aircraft was made in December 1903 by the Wright brothers, some sixty-five years later – less than a man's lifetime – Neil Armstrong stepped on the moon in July 1969.

Certainly, technological progress comes with its own set of risks, as has always been the case. However, it's crucial also to consider the benefits. In the post-"fourth wave" future, we can expect smarter apps capable of explaining their actions and understanding ours. This advancement will not only deepen our self-understanding but also improve our understanding of others, ultimately enhancing our quality of life.

Artificial Intelligence can improve the quality of services and/or products by introducing previously unattainable features. For example, Google's competitive edge in online search is partly due to intelligent software and database functions that deliver highly relevant results swiftly, sometimes even providing direct answers to user queries. Systems like IBM's Watson assist doctors in diagnosing patients by drawing upon extensive medical literature. Another prime example is the recommendation engines employed by Netflix and Amazon, which suggest books and movies tailored to individual customers (Vrabie, 2022).

As it should be clear by now, virtually all practical applications of AI involve a human-computer partnership. Humans primarily engage in software creation and maintenance. They are also the decision-makers when it comes to selecting applications and troubleshooting machine failures. Most crucially, humans are indispen-sable for performing tasks that computers are currently incapable of handling on their own.

Therefore, the systems we aim to optimize and design are not merely computer systems in the strictest sense but human-machine systems. To achieve this, we first need to address two questions. The first is, "Which tasks should be performed by computers, and which by humans?" The second is, "How can we improve this human-machine system over time?"

Starting with the first question, an ideal approach is to allocate tasks based on each entity's strengths: let machines handle what they excel at and let humans do what they are better at. For instance, machines excel at storing and recalling vast amounts of information, while humans generally have a more remarkable aptitude for social interactions. The goal shouldn't be to replace one with the other wherever possible but to determine how these symbiotic systems can perform tasks more effectively than any individual or computer – or even groups of individuals or networks of computers – could achieve independently.

Let's take Google's search engine as an example. Depending on some perspectives, it could be considered the most widely used IT/AI system today. In this system, humans primarily create and link content, while machines store enormous amounts of information, aiding users in finding the content most relevant to their needs. While the Google search engine has in some ways supplanted librarians, who performed a somewhat similar function in the past, it has also democratized content creation and facilitated more exhaustive searches in the vast online library of knowledge. Although one could argue that Google has contributed to the decline of the librarian profession, it has also created numerous new jobs that involve both searching for relevant information and generating new content, as well as advertising it to users – consider YouTube as an example, a platform that helps millions to build a well-rewarded business in various domains.

Wikipedia is another well-known and successful case. While humans are still the primary creators and editors of content, machine algorithms, like Wikipedia's software bots, excel at quickly scanning for issues such as inappropriate language or plagiarism. This symbiotic relationship between humans and machines is more effective than either working independently. Wikipedia has not only supplanted printed encyclopedias to some extent but has also significantly broadened the scope and timeliness of the information available, becoming our go-to source for quick and, to some extent, reliable information.

Cybersecurity tools like antivirus software and firewalls are further examples of effective human-machine collaboration. In these applications, machines excel at detecting unusual network activity, while humans are better at discerning which types of "unusual" activity are actually legitimate actions. In the realm of cybersecurity, collaborative efforts between humans and machines have proven to be highly effective, detecting up to three times more malicious attacks than electronic systems operating independently (Almahmoud, Yoo, Alhussein, Farhat, & Damiani, 2023; Sommer & Paxson, 2010).

## (Few) "roles" of AI

To better understand the division of tasks between humans and computers, it's helpful to consider the roles computers can play in relation to us.

Firstly, computers often serve as tools, performing tasks under close human supervision. Examples include word processors with auto-completion and correction features, spreadsheets, and cars equipped with cruise control. Most computer applications today serve as tools or platforms that connect people to other people or machines (Schachtner, 2021). Consider email, the World Wide Web, Netflix, and Facebook as instances of this. Increasingly, apps are being developed to facilitate human communication through innovative collaborative features. This phenomenon, often termed "hyper-connectivity," represents a significant way information technology contributes to creating smarter institutions and organizations.

Another avenue for creating smarter institutions and organizations lies in developing more intelligent machines. These machines should benefit both the service recipients and the staff of institutions employing AI technologies. A prime example of this dual role comes from Google. As mentioned above, its search algorithms serve as tools, delivering results almost instantaneously. However, another component of Google's technology continuously scans and indexes the web, functioning more as an assistant than a mere tool. Unlike tools, assistants can operate without the user's direct attention and proactively assist in problem-solving. They may even initiate operations independently when triggered – consider semi-autonomous cars as a form of driver's assistance.

IBM Watson technology, aptly named after Sherlock Holmes' assistant, offers another illustrative example. It has ingested vast amounts of medical literature and is increasingly aiding doctors globally in diagnosing various medical conditions. This success hinges on an interactive dialogue between the doctor and the application. The doctor provides specific case information, and the application employs mathematical reasoning – based on statistics and probabilities – to generate possible diagnoses and the logical pathways leading to them. These analytical results are then presented to the doctors, who can probe further if needed and ultimately make the final decisions on diagnosis and treatment.

Other compelling examples of assistants are found in industries that previously relied heavily on large call center departments, such as mobile operators and airlines. Nowadays, robots or chatbots handle customer queries and issues through social networks or even phone calls. Artificial intelligence software generates suggested responses to customer messages. However, these suggestions are often reviewed by human agents who may choose to override the machine-generated response with their own.

In some cases, machines can function in a role similar to that of humans, essentially as equals. In these scenarios, machines perform tasks that closely resemble those carried out by humans. Take the insurance company Lemonade as an example (Lemonate, 2015). Customers can use mobile apps to file insurance claims, and if a claim meets certain predefined criteria, the AI system can automatically approve and pay out the claim within seconds. However, the claim is rerouted to a human operator for review in atypical situations.

Another striking example of a computer functioning in a role akin to humans is in the realm of chess. AlphaZero, an application we discussed earlier, serves as a personal trainer for the former world chess champion, Magnus Carlsen (AlphaZero, 2019). A few years ago, intricate moves in a game would be analyzed post-match by an entire team aiming to uncover better or defensive moves, thereby making various predictions. Today, analysis engines offer insights of a depth unparalleled by human cognition. These engines provide strategic game plans designed to catch opponents off guard[15]. Notably, these games are played without real-time computer assistance; the game's dynamics rely on the players' comprehension of the machine's advice and their ability to remember it (Silver et al., 2018).

There have also been "centaur" experiments in chess, named after the mythical creatures from Greek mythology that combine a horse's body with a human torso, symbolizing the union of human intelligence and animal strength. These experiments, approved by the International Chess Federation (FIDE), involve top world players who are live-assisted by computers during the gameplay.

Here are a few instances where computers act in roles traditionally filled only by humans – or even entire teams of strategic analysts and tacticians. Additional examples can be found in stock market trading, where software robots execute buy/sell orders based on deep market analysis. Recently, social media platforms like YouTube have been scanned by software bots for keywords. These bots collect, edit, and aggregate videos with similar content – such as nature, travel, or humor – and present them to users for viewing, all without human intervention. The effectiveness of such a software bot directly correlates with its ability to generate views.

Lastly, computers can also serve in a managerial role in relation to humans (Boce, Tomço, & Hyra, 2022). While the idea of a machine acting as a manager may raise concerns for some, this is not a new concept – consider traffic lights, which essentially manage traffic flow at intersections, effectively replacing human traffic police. Most people find this automated management entirely acceptable.

---

[15] During the Norway Chess Open, Magnus Carlsen defeated American Wesley So with an amazing concept that AlphaZero had found in a training game.

Certainly, there are many scenarios where computer systems, if given the authority to manage humans, would do so in entirely unacceptable ways. However, the focus here is on systems that can effectively take on tasks traditionally reserved for humans.

Another excellent example of managing intellectual capital is the research system CrowdForge, developed by Carnegie Mellon University in the United States (Carnegie Mellon University, 2023; Kittur, Smus, Khamkar, & Kraut, 2011). Researchers used this system to coordinate a group of people in writing encyclopedia articles, and one of the most successful articles was about New York City. To accomplish this, they enlisted individuals from Amazon Mechanical Turk's online job marketplace, where participants engage in "micro-tasks" (small tasks often lasting just a few minutes and paying only a few cents). For the New York City article, the researchers first asked a group to create a table of contents that included sections on sights, history, culture, and so on. Subsequently, they tasked another group to generate or find relevant facts for each section in the table of contents. These facts were then passed on to different individuals who crafted coherent paragraphs from them. Finally, the computer system assembled these paragraphs into a complete article. In this scenario, humans served primarily as executors of specific tasks, aside from those handling the technical aspects required to keep the computer system running smoothly. When independent observers were asked to evaluate these articles, they found them to be of higher quality than those written by a single individual for the same cost. The articles were also deemed roughly equivalent in quality to Wikipedia articles.

In this project, humans executed all the tasks, while the machine merely coordinated the workflow, delegating tasks to different teams. While the machine's role in this project may not have been particularly complex or reliant on advanced artificial intelligence, researchers are now developing more sophisticated methods for task assignment. These methods enable machines to decide who should perform specific tasks based on past performance in similar roles.

Starting from this successful example, we can envision a future where artificial intelligence systems excel at managing workflow and task allocation within institutions. Such systems could not only streamline internal operations but also facilitate interdepartmental collaboration.

Another managerial role often involves staff appraisal. One example is the Cogito system (Cogito, 2023). Cogito is designed to help phone representatives enhance their conversations by providing real-time feedback on the emotional quality of their interactions. While the system doesn't collect enough data to continue the conversation independently, it listens for subtle cues like tone of voice and response time. This allows it to offer valuable feedback to phone operators, helping them improve their empathy and communication skills.

So far, we've primarily discussed how tasks can be divided between humans and computers. The other crucial question we need to address is how this human-computer system can improve in the future. A few decades ago, the business sector – and not just it – was keenly focused on analyzing all processes to make them more efficient. This led to a shift in investment toward automation, which gradually became the primary goal for many companies.

Public institutions soon followed the lead of their private counterparts, overhauling their entire information flow to better meet citizens' demands. However, artificial intelligence goes beyond mere automation. As software applications continue to improve, often learning from their own experiences, the line between tasks that can only be performed by humans and those that can be automated will increasingly blur (Vrabie, 2023).

This perspective aligns with the views of Andy Clark, who not only outlines a future landscape but also encourages us to see the human-computer partnership as an ever-evolving system. This system learns from experience to become increasingly efficient (Clark, 2003). There are at least three avenues through which this can be successfully achieved.

- Firstly, there's the human component. We need to figure out how to improve upon what we're already doing. To illustrate this point, we should focus on streamlining processes by identifying and eliminating redundant steps.
- A second approach involves programmers continually refining machine capabilities. For instance, as are the developers at Meta/Facebook and other tech companies, Google's engineers are always tweaking and improving search algorithms.
- A third avenue is through machine self-improvement using various machine learning models. I previously mentioned chatbot applications, supervised by humans to sometimes correct the AI's suggested responses. When this occurs, the AI system learns from the experience, becoming more adept at providing a human-like answer in similar future scenarios (Vrabie, 2023).

The realm of artificial Intelligence is in a constant state of evolution, with advancements sometimes occurring so rapidly that they astonish even the most seasoned experts in the field. There's already a push to integrate AI into every aspect of our lives, making the potential for progress virtually limitless. Consider Google Translate as an example; it has evolved from a basic text-based tool to a sophisticated system capable of facilitating real-time voice conversations between users who don't share a common language. Google Duplex, also known as Google Assistant, is another impressive system that can conduct phone conversations with people and handle tasks like making restaurant reservations or scheduling different appointments.

Research and development in autonomous vehicles (AVs) continue to amaze automotive and technology enthusiasts. Initially, these vehicles were designed to follow well-mapped routes and struggled with unexpected situations, especially in busy urban areas. Nowadays, they can change lanes and overtake other cars in a manner that mimics human behavior. Additionally, the MapLine app has been developed to assist AVs in navigating roads that haven't been electronically mapped before. The future of autonomous vehicles is becoming a distinct subfield within artificial Intelligence.

**AI in education**

In a recent study (Vrabie, 2023), following a project entitled "Digital Media for Enhancing Educational Quality and Facilitating Labor Market Entry: Leveraging Visual Technologies in Virtual Pedagogical Environments – E-qual-EDU" led by a team of researchers from Smart-EDU Hub within the National University of Political

Studies and Public Administration (SNSPA) (Smart-EDU Hub, 2021), the author substantiated through empirical analysis the pivotal role of Artificial Intelligence (AI) and gamification mechanisms in increasing student engagement and facilitating knowledge retention. Despite specific reservations and limitations, the study finds that educational frameworks are poised to integrate AI and gamification methodologies (Vrabie, 2023), thereby gaining increasing momentum in the forthcoming years (Patruti, Zbuchea, & Pînzaru, 2023).

The research delineates the nuanced distinctions between the concepts of digitization and digitalization, accentuating that the latter facilitates active user engagement and capitalizes on AI methodologies for a more dynamic transformation of content (Tarziu & Vrabie, 2015). As academia transitions into the paradigm of Education 3.0, higher education institutions confront a critical inflection point. Swift acclimatization to emergent technologies is imperative for student recruitment, notwithstanding the disparate levels of student commitment and extant societal limitations. The literature review and the aforementioned research project corroborate that contemporary e-learning platforms substantially amplify student performance by enhancing both the retention and understanding of information (Iancu, Vrabie, & Ungureanu, 2021). Within this context, the potential for AI to serve as pedagogical adjuncts can also be contemplated.

Moreover, one can envisage an AI-driven pedagogical agent capable of "translating" instructional materials and autonomously generating assessments – tailored to individual psychographic profiles, such as specific interests (e.g., athletics), cognitive preferences (e.g., visual learning modalities) and personality traits (e.g., extroversion). Such personalized interventions could invigorate student engagement and stimulate a more profound learning experience (Krumova, 2017). During the learning process, students can ask a virtual assistant – enriched with chat capabilities such as GPT-4, relevant questions on the subject in focus. The "assistant" can reply by elaborating on the course material or even going beyond it and searching for relevant information on designated bibliographic resources.

Additionally, in the realm of career development, it is often observed that younger individuals frequently lack a definitive vision for their vocational trajectory. One could hypothesize the deployment of an AI-based career advisory system capable of discerning each student's hobbies, interests, innate talents, and coping mechanisms. Utilizing this data, the system could facilitate the exploration of bespoke career pathways, drawing upon the vocational experiences of analogous professionals. Such an AI career advisory system could further guide students in optimally seizing these vocational opportunities, thereby maximizing their career potential.

**Takeaways**

To harness the transformative potential of AI in the educational domain, it is imperative to adhere to a set of guiding principles:
- The AI system must be endowed with a comprehensive and accurate knowledge base sourced from academically rigorous course materials.
- For AI to generate personalized life scenarios, it necessitates an in-depth understanding of user psychographics alongside other user-specific attributes.

- Incorporating human expertise is essential for rectifying algorithmic inaccuracies or errors, thereby serving as a corrective mechanism within the AI system.

We stand at the cusp of a nascent era in AI, one that has already expanded the horizons of what was previously conceivable. While the liberal application of this technology is not without its risks—evidenced by the evolution of platforms like Facebook from collegiate social networks to conduits for election interference and extremist propaganda—these unintended consequences should not overshadow the manifold benefits. This article's author contends that society will adapt to the challenges posed by generative AI, much like other technological and industrial innovations that also carry inherent risks.

**AI: beyond computing**

The discourse surrounding artificial Intelligence (AI) has elicited a spectrum of perspectives among experts in the field. While there is a general accord regarding AI's transformative potential within the computing industry, its broader implications – spanning economic, legal, and military domains – remain enveloped in a veil of uncertainty. This ambiguity complicates not only the understanding of AI's current impact but also the forecasting of its future trajectory. In this context, the present paper's efforts were to investigate the strategic deployment of AI technologies to increase societal well-being.

By adopting a multi-disciplinary approach, we aim to contribute to the ongoing dialogue and offer actionable insights for the responsible integration of AI into various sectors of society.

**Acknowledgment**
It is important to note that segments of this article are slated for inclusion in an upcoming publication entitled "Artificial Intelligence from Idea to Implementation." This cross-publication aims to disseminate the research findings to a broader academic audience and contribute to the evolving discourse on the role of Artificial Intelligence in today's society.

# References

Almahmoud, Z., Yoo, P., Alhussein, O., Farhat, I., & Damiani, E. (2023). A holistic and proactive approach to forecasting cyber threats. *Nature, 13*. https://doi.org/10.1038/s41598–023-35198–1

AlphaZero. (2019). *The exciting impact of a game changer.* Retrieved from https://www.newinchess.com/media/wysiwyg/product_pdf/872.pdf

Baltac, V. (2017). *Mituri și realitate în lumea digitală [Myths and reality in the digital world].* EXCEL XXI Books.

Barr, A., & Feigenbaum, E. A. (Eds.). (1981). *The Handbook of Artificial Intelligence.* Butterworth-Heinemann.

Boce, G., Tomço, V., & Hyra, A. (2022). IT tools and strategies in business. *Smart Cities and Regional Development (SCRD), 6*(1), 79–89.

Campbell, C. (2023). *What is 'Smart' Technology?* (Williams College) Retrieved from https://oit.williams.edu/ats-posts/what-is-smart-technology/

Carnegie Mellon University. (2023). *CrowdForge*. (Carnegie Mellon University) Retrieved from https://crowdforge.net/

Clark, A. (2003). *Natural-Born Cyborgs: Minds, Technologies, and the Future of Human Intelligence.* Oxford University Press.

Cogito. (2023). *Powering the Teams Behind Great CX & EX*. (Cogito) Retrieved from https://cogitocorp.com/

Fast Company. (2023, April 9). *Why Meta's Yann LeCun isn't buying the AI doomer narrative*. (Fast Company) Retrieved

from https://www.fastcompany.com/90947634/why-metas-yann-lecun-isnt-buying-the-ai-doomer-narrative

Gardner, H. (1983). *Frames of Mind: A Theory of Multiple Intelligences.* Basic Books.

Garry, K., & Mig, G. (2017). *Deep Thinking: Where Machine Intelligence Ends and Human Creativity Begins.* PublicAffairs.

Goody, J., & Bell, D. (1975). *Literacy in Traditional Societies.* Cambridge University Press.

Unruh, G., & Kiron, D. (2017, November 6). *Digital Transformation on Purpose.* Retrieved from https://sloanreview.mit.edu/article/digital-transformation-on-purpose/

Iancu, D., Vrabie, C., & Ungureanu, M. (2021). Is Blended Learning Here to Stay? Public Administration Education in Romania. *Central and Eastern European eDem and eGov Days.* Budapest. https://doi.org/10.24989/ocg.v341.4

IEEE Spectrum. (2023, April 19). *GPT-4, AGI, and the Hunt for Superintelligence*. (IEEE Spectrum) Retrieved from https://spectrum.ieee.org/superintelligence-christoph-koch-gpt4

Image net. (2012). *Large Scale Visual Recognition Challenge 2012 (ILSVRC2012)*. Retrieved from https://image-net.org/challenges/LSVRC/2012/results.html

JordanTeslaTech. (2021, July 23). *Twitter.* Retrieved from https://twitter.com/jordanteslatech/status/1418413307862585344?lang=ro

Kittur, A., Smus, B., Khamkar, S., & Kraut, R. (2011). CrowdForge: Crowdsourcing Complex Work. *ACM*. Retrieved from https://static.googleusercontent.com/media/research.google.com/ro//pubs/archive/39980.pdf

Krumova, M. (2017). Open Data Benchmarking for Higher Education: Management and Technology Perspectives. *Smart Cities and Regional Development (SCRD) Journal, 1*(2), 47–60. Retrieved from https://www.scrd.eu/index.php/scrd/article/view/17

Lazar, M. A., Zbuchea, A., & Pînzaru, F. (2023). The Emerging Generation Z Workforce in the Digital World: A Lit-

erature Review on Cooperation and Transformation. *Proceedings of the International Conference on Business Excellence, 17*(1), 1991–2001. https://doi.org/10.2478/picbe-2023–0175

LeCun, Y. (2020). AI, Deep Learning, and Machine Learning.

Lemonate. (2015). *Lemonade: An Insurance Company Built for the 21st Century*. (Lemonade) Retrieved from https://www.lemonade.com/

McCarthy, J., Minsky, M., Rochester, N., & Shannon, C. (1955, August 31). *Stanford.edu.* Retrieved from http://jmc.stanford.edu/articles/dartmouth/dartmouth.pdf

Minsky, M. (1960). Steps Toward Artificial Intelligence. *Proceedings of the IRE, 49*(1), 8–30. https://doi.org/10.1109/JRPROC.1961.287775

Minsky, M. (2006). *The Emotion Machine: Commonsense Thinking, Artificial Intelligence, and the Future of the Human Mind.* Simon & Schuster Paperbacks.

NanoWerk. (2023). *What Is Smart Technology?* (NanoWerk) Retrieved from https://www.nanowerk.com/smart/what-is-smart-technology.php

OpenAI. (2023, March 31). *GPT-4 is not aware of its existence – works like GPT-3.5*. (OpenAI) Retrieved from https://community.openai.com/t/gpt-4-is-not-aware-of-its-existence-works-like-gpt-3–5/132179

Patruti, P., Zbuchea, A., & Pînzaru, F. (2023). Fashion Joining Online Gaming and the Metaverse. *Proceedings of the International Conference on Business Excellence, 17*(1), 1065–1074. https://doi.org/10.2478/picbe-2023–0096

Schachtner, C. (2021). Smart government in local adoption – Authorities in strategic change through AI. *Smart Cities and Regional Development (SCRD) Journal, 5*(3), 53–61.

Silver, D., et al. (2018). Mastering Chess and Shogi by Self-Play with a General Reinforcement Learning Algorithm. *Science, 362*(6419), 1140–1144.

Slagle, J. (1961). *A heuristic program that solves symbolic integration problems in freshman calculus, Symbolic Automatic*

*Integrator (SAINT).* Retrieved from https://dspace.mit.edu/bitstream/handle/1721.1/11997/31225400-MIT.pdf?sequence=2

Smart-EDU Hub. (2021, December). *e-QUAL EDU*. (Smart-EDU Hub, SNSPA) Retrieved from https://www.smart-edu-hub.eu/events/projects

Sommer, R., & Paxson, V. (2010). Outside the Closed World: On Using Machine Learning for Network Intrusion Detection. *2010 IEEE Symposium on Security and Privacy.* Retrieved from https://www.computer.org/csdl/proceedings-article/sp/2010/05504793/12OmNA0dMRJ

Tarziu, A., & Vrabie, C. (2015). *Education 2.0: universities' e-learning methods.* LAP Lambert Academic Publishing.

The New York Times. (1981, August 13). Big I.B.M.'s Little Computer. *The New York Times.* Retrieved from https://www.nytimes.com/1981/08/13/business/big-ibm-s-little-computer.html

The New York Times. (2023, March 15). GPT-4 Is Exciting and Scary. *The New York Times.* Retrieved from https://www.nytimes.com/2023/03/15/technology/gpt-4-artificial-intelligence-openai.html

Turing, A. (1950). Computing Machinery and Intelligence. *Mind, 49*, 433–460.

UPB. (2021, March 21). *MECIPT-1 – Electronic Computing Machine of the Timişoara Polytechnic Institute*. (UPB) Retrieved from https://www.upt.ro/Informatii-utile_mecipt-1---electronic-computing-machine-of-the-timisoara-pol_618_en.html

Vidu, C., Zbuchea, A., & Pinzaru, F. (2021). Old Meets New: Integrating Artificial Intelligence in Museums' Management Practices. In C. Bratianu et al. (Eds.), *Strategica. Shaping the Future of Business and Economy* (pp. 830–844), Tritonic. Retrieved from https://strategica-conference.ro/wp-content/uploads/2022/04/63–1.pdf

Vidu, C., Zbuchea, A., Mocanu, R., & Pinzaru, F. (2020). Artificial Intelligence and the Ethical Use of Knowledge. In C. Bra-

tianu et al. (Eds.), *Strategica. Preparing for Tomorrow, Today* (pp. 773–784), Tritonic.

Vrabie, C. (2016). *Elements of e-government.* Bucharest: Pro Universitaria.

Vrabie, C. (2022, December 8). Artificial Intelligence Promises to Public Organizations and Smart Cities. *Digital Transformation. Lecture Notes in Business Information Processing, 465*. https://doi.org/10.1007/978–3-031–23012-7_1

Vrabie, C. (2023). Education 3.0 – AI and Gamification Tools for Increasing Student Engagement and Knowledge Retention. *Digital Transformation. Lecture Notes in Business Information Processing, 495*, 74–87. https://doi.org/10.1007/978–3-031–43590-4_5

Vrabie, C. (2023). E-Government 3.0: An AI Model to Use for Enhanced Local Democracies. *Sustainability*. https://doi.org/10.3390/su15129572

Vrabie, C., & Eduard, D. (2018). *Smart cities: de la idee la implementare, sau, despre cum tehnologia poate da strălucire mediului urban.* Universitara.

Winston, P. H. (1990). *Oral history interview with Patrick H. Winston.* Charles Babbage Institute.

Yalalov, D. (2023, March 22). *GPT-4 Tests Indicate That It Already 'Conscious'*. Retrieved from https://mpost.io/gpt-4-tests-indicate-that-it-already-conscious/